\documentclass[twocolumn,showpacs,aps,floatfix,prd]{revtex4}

\usepackage{graphicx}
\usepackage{dcolumn}
\usepackage{amsmath}
\usepackage{epsfig}

\input babarsym.tex

\newcommand{\BABARPubYear}    {07}
\newcommand{\BABARPubNumber}  {015}

\newcommand{\SLACPubNumber} {12417}

\def\figurebox#1#2#3{%
    \def\arg{#3}%
    \ifx\arg\empty
    {\hfill\vbox{\hsize#2\hrule\hbox to #2{\vrule\hfill\vbox to #1{\hsize#2\vfill}\vrule}\hrule}\hfill}%
    \else
    {\hfill\epsfbox{#3}\hfill}%
    \fi}

\newcommand{\Do}{D^0}

\newcommand{\GeV}{\rm{GeV}}
\newcommand{\GeVc}{\rm{GeV/c}}
\newcommand{\GeVcd}{\rm{GeV/c^2}}
\newcommand{\MeV}{\rm{MeV}}

\newcommand{\MeVcd}{\rm{MeV/c^2}}

\newcommand{\ba}{\begin{array}}
\newcommand{\ea}{\end{array}}
\newcommand{\bc}{\begin{center}}
\newcommand{\ec}{\end{center}}
\newcommand{\beq}{\begin{eqnarray}}
\newcommand{\eeq}{\end{eqnarray}}
\newcommand{\bes}{\begin{eqnarray*}}
\newcommand{\ees}{\end{eqnarray*}}
\newcommand{\Zz}{\ifmmode {\rm Z} \else ${\rm Z } $ \fi}
\newcommand{\xxbar}{\ifmmode {\rm x\bar{x}} \else ${\rm x\bar{x}} $ \fi}
\newcommand{\rphi}{\ifmmode {\rm R\phi} \else ${\rm R\phi} $ \fi}

\long\def\inst#1{\par\nobreak\kern 4pt\nobreak
  {\it #1}\par\vskip 10pt plus 3pt minus 3pt}

\begin{document}


\begin{flushleft}
\babar-PUB-\BABARPubYear/\BABARPubNumber\\
SLAC-PUB-\SLACPubNumber\\
[10mm]
\end{flushleft}

\title{
{\large \bf
Measurement of the  Hadronic Form Factor in 
\boldmath{$\Do \rightarrow \Km e^+ \nu_e $} Decays. } 
}

%
\author{B.~Aubert}
\author{M.~Bona}
\author{D.~Boutigny}
\author{Y.~Karyotakis}
\author{J.~P.~Lees}
\author{V.~Poireau}
\author{X.~Prudent}
\author{V.~Tisserand}
\author{A.~Zghiche}
\affiliation{Laboratoire de Physique des Particules, IN2P3/CNRS et Universit\'e de Savoie, F-74941 Annecy-Le-Vieux, France }
\author{J.~Garra~Tico}
\author{E.~Grauges}
\affiliation{Universitat de Barcelona, Facultat de Fisica, Departament ECM, E-08028 Barcelona, Spain }
% ...
\collaboration{The \babar\ Collaboration}
\noaffiliation

\date{\today}

\begin{abstract}

The shape of the hadronic form factor $f_+(q^2)$ in the decay $\Do \rightarrow \Km e^+ \nu_e$
has been measured in a model independent analysis and compared with theoretical calculations. 
We use 75$\fb^{-1}$ of data recorded by the \babar\ detector at the PEPII electron-positron collider.
The corresponding decay branching fraction, relative to the decay $D^0 \rightarrow \Km \pi^+$, 
has also been measured to be 
$R_D = BR(D^0 \rightarrow K^- e^+ \nu_e)/BR(D^0 \rightarrow K^- \pi^+) = 0.927 \pm 0.007 \pm 0.012$.
From these results, and using the present world average value for
$BR(D^0 \rightarrow K^- \pi^+)$,
the normalization of the form factor at $q^2=0$ is determined to be 
$f_+(0)=0.727\pm0.007\pm0.005\pm0.007$
where the uncertainties are statistical, systematic, and from external inputs, 
respectively.

\end{abstract}


\maketitle


\section{Introduction}
\label{sec:Introduction}
Measurements of exclusive semileptonic $D$ decays provide an accurate 
determination of the hadronic form factors entering in these decays. Assuming that the
CKM matrix is unitary, the elements $\left | V_{cs} \right |$ and  $\left | V_{cd} \right |$ 
can be determined:
\beq
 \left |V_{cs}\right | = 
\left | V_{ud}\right | -\frac{\left | V_{cb}\right |^2 }{2} +{\cal O}(\lambda^6) = 0.9729 \pm 0.0003,
\eeq
using the measured values  \cite{ref:pdg06} of $\left | V_{ud}\right |$ and $\left | V_{cb}\right |$, and
the sine of the Cabibbo angle $\lambda=\sin(\theta_c)\simeq 0.227$. 
Theoretical predictions give estimates of the form factors in exclusive semileptonic $B$
 and $D$ meson decays. Precise measurements of the hadronic form factors in $D$ decays can help 
to validate predictions from QCD calculations in both $D$ and $B$ decays. Better understanding of 
the form factors in $B$ decays is necessary to improve the precision on the determination of 
$\left | V_{cb} \right |$ and  $\left | V_{ub} \right |$.

In  
$\Do \rightarrow \Km e^+ \nu_e$ decays \cite{ref:chargeconj}, 
with a pseudoscalar hadron emitted
in the final state, and neglecting the electron mass,
the differential decay rate depends only on one form factor $f_+(q^2)$,
\beq
 \frac{d \Gamma}{d q^2} = \frac{G^2_F}{24 \pi^3} \left | V_{cs} \right |^2
\left | \vec{p}_K (q^2) \right |^3 \left |f_+(q^2) \right |^2,
\eeq
where $G_F$ is the Fermi constant, 
$q^2$ is the invariant mass squared of the two leptons, $e^+$ and $\nu_e$, and 
$\vec{p}_K (q^2)$ is the kaon three-momentum in the $D^0$ rest frame \cite{ref:kin}.
In this paper we present measurements of the $q^2$ variation and absolute
value of the hadronic form factor at $q^2 = 0$ for the
decay $\Do \rightarrow \Km e^+ \nu_e(\gamma)$.  The data consist of $D$ 
mesons produced in $e^+e^- \to c\bar{c}$ continuum events
at a center of mass energy near the $\Upsilon(4S)$ mass, and 
were recorded by the \babar\ detector at the Stanford 
Linear Accelerator Center's PEP-II collider.
A semi-inclusive reconstruction technique is used to select
charm semileptonic decays with high efficiency. As a result of this approach,
events with a photon radiated during the $D^0$ decay are included in the signal.  
The systematic uncertainties are kept as low as possible by using control samples
extracted from data where possible. 

Measurements of $D \rightarrow K \bar{\ell} \nu_{\ell}$, based on smaller signal events samples, have been 
published by the CLEO \cite{ref:cleo3}, FOCUS \cite{ref:focus} and
Belle \cite{ref:belle} collaborations.

This paper is organized as follows. A general description of the hadronic form factor, $f_+(q^2)$,
is given in Section \ref{sec:fq2}, where the different parameterizations 
considered in this analysis are
explained. In Section \ref{sec:babar}  a short description of the detector components
that are important to this measurement is given. The selection of signal
events and the rejection of background are considered in Section \ref{sec:Analysis}.
In Section \ref{sec:Physics}, the measured $q^2$ variation of the hadronic form factor 
is discussed and compared with previous measurements. 
In Section \ref{sec:rate} the measured decay rate is given
and in Section \ref{sec:Summary} these measurements are combined to
obtain the value of $f_+(0)$.

\section{The \boldmath{$f_+(q^2)$} hadronic form factor}
\label{sec:fq2}
The amplitude for the decay $\Do \rightarrow \Km \ell^+ \nu_{\ell}$ depends on two hadronic form factors:
\beq 
<K(p^{\prime})|V_{\mu}|D(p)> &= &
\left ( p_{\mu}+p^{\prime}_{\mu}-q_{\mu} \frac{m_D^2-m_K^2}{q^2} \right )f_+(q^2)\nonumber\\
&+&\frac{m_D^2-m_K^2}{q^2} q_{\mu}f_0(q^2)  
\eeq
where $V_{\mu}=\bar{s} \gamma_{\mu}c$. 
The constraint $f_+(0)=f_0(0)$ ensures that there is no singularity at $q^2=0$.
When the charged lepton
is an electron, 
the contribution from $f_0$ is proportional to $m_e^2$ and can be neglected 
in decay rate measurements.

The parameterizations of $f_+(q^2)$ which have been compared with present
measurements and a few examples of 
theoretical approaches, proposed to determine the values of corresponding
parameters, are considered in the following.  

\subsection{Form factor parameterizations}
The most general expressions of the form factor $f_+(q^2)$ are analytic
functions satisfying the dispersion relation:
\beq
f_+(q^2) = \frac{Res (f_+)_{q^2=m^2_{D_s^*}}}{ m_{D_s^*}^2-q^2}
+ \frac{1}{\pi} \int_{t_+}^{\infty} dt \frac{\Im{f_+(t)}}{t-q^2-i\epsilon}.
\eeq
The only singularities in the  complex $t\equiv q^2$ plane
originate from the interaction of the charm and the strange quarks
in vector states. They are a pole, situated at the $D_s^*$ mass squared
and a cut, along the positive real axis, starting at threshold ($t_+=(m_D+m_K)^2$) 
for $D^0K^-$ production.

\subsubsection{Taylor expansion}
This cut $t$-plane can be mapped onto the open unit disk with center at $t=t_0$
using the variable:
\beq
z(t,t_0) = \frac{\sqrt{t_+-t}-\sqrt{t_+-t_0}}{\sqrt{t_+-t}+\sqrt{t_+-t_0}}.
\eeq 
In this variable, the physical region for the semileptonic decay ($0<t<t_-=q^2_{\rm max}=(m_D-m_K)^2$)
corresponds to a real segment extending between $\pm z_{\rm max}=\pm 0.051$. This value of $z_{\rm max}$ is obtained  for
$t_0 =t_+ \left ( 1 -\sqrt{1-t_-/t_+}\right )$.
The $z$ expansion of $f_+$ is thus expected to converge quickly.  
The most general parameterization \cite{ref:hill1}, consistent with constraints from QCD, 
\beq
f_+(t)=\frac{1}{P(t) \Phi(t,t_0)}\sum_{k=0}^{\infty}a_k(t_0)~ z^k(t,t_0),
\label{eq:taylor}
\eeq
is  based on earlier considerations \cite{ref:beforehill}. The function
$P(t)=z(t,m_{D_s^*}^2)$ has a zero at the $D_s^*$ pole mass and $|P|=1$ along the unit circle;
$\Phi$ is given by:

\beq
\Phi(t,t_0) &=& \sqrt{\frac{1}{24 \pi \chi_V}}
\left ( \frac{t_+-t}{t_+-t_0}\right )^{\frac{1}{4}}
\left ( \sqrt{t_+-t}+\sqrt{t_+} \right )^{-5}\nonumber\\
&\times&\left ( \sqrt{t_+-t}+\sqrt{t_+-t_0} \right )\\
&\times&\left ( \sqrt{t_+-t}+\sqrt{t_+-t_-} \right )^{\frac{3}{2}}\left ( t_+-t \right)^{\frac{3}{4}}\nonumber,
\eeq

\noindent 
where $\chi_V$ can be obtained from dispersion relations using perturbative
QCD and depends on $u=m_s/m_c$ \cite{ref:lebed}. 
At leading order, with $u=0$ \cite{ref:chiu},
\beq
\chi_V = \frac{3}{32 \pi^2 m_c^2}.
\eeq
\noindent
The choice of $P$ and $\Phi$ is such that:
\beq
\sum_{k=0}^{\infty}a_k^2(t_0)\leq 1 \label{eq:unitary}.
\eeq
Having measured the first coefficients of this expansion, 
Eq. (\ref{eq:unitary})
can constrain the others. This constraint, which depends
on  $\chi_V$, may have to be abandoned in the case
of charm decays as the charm-quark mass may not be large enough
to prevent the previous evaluation
of $\chi_V$ from receiving large $1/m_c$ and QCD corrections. However the parameterization given
in Eq. (\ref{eq:taylor}) remains valid and
it has been compared \cite{ref:hill1} with available measurements.
The first
two terms in the expansion were sufficient to describe the data.

\subsubsection{Model-dependent parameterizations}
A less general approach assumes that the $q^2$ variation of 
$f_+(q^2)$ is
governed mainly by the $D_s^*$ pole and that the other contributions can be accounted for
by adding another effective pole at a higher mass \cite{ref:bk}:

\beq
f_+(q^2) &=&\frac{f_+(0)}{1-\alpha_{\rm pole}} \left ( \frac{1}{1-\frac{q^2}{m_{D_s^*}^2}} 
-\frac{\alpha_{\rm pole}}{1-\frac{q^2}{\gamma_{\rm pole} m_{D_s^*}^2}}
\right )\nonumber\\
&=& f_+(0) \frac{1-\delta_{\rm pole}\frac{q^2}{m_{D_s^*}^2}}{\left (1-\frac{q^2}{m_{D_s^*}^2}\right ) \left (1-\beta_{\rm pole} \frac{q^2}{m_{D_s^*}^2} \right )}
\label{eq:twopoles}
\eeq
with $\delta_{\rm pole} = (1/\gamma_{\rm pole} -\alpha_{\rm pole})/(1 -\alpha_{\rm pole})$ and $\beta_{\rm pole}=1/\gamma_{\rm pole}$.

If in addition, the form factors $f_+$ and $f_0$ must obey
a relation, valid at large recoil and in the heavy quark limit,
then $\alpha_{\rm pole} = 1/\gamma_{\rm pole}$ \cite{ref:bk}
($\beta_{\rm pole} = \alpha_{\rm pole}$ and $\delta_{\rm pole}=0$ in this case). Equation 
(\ref{eq:twopoles}) becomes:

\beq
f_+(q^2) =  \frac{f_+(0)}{ \left (1-\frac{q^2}{m_{D_s^*}^2}\right ) 
\left ( 1-\alpha_{\rm pole} \frac{q^2}{ m_{D_s^*}^2}\right )},
\label{eq:modpolemass}
\eeq
known as the modified pole ansatz.
Initially an even simpler expression, the simple pole ansatz, was proposed
which considered
only the contribution from the $D_s^*$ pole.
In the following, the pole mass entering
in
\beq
f_+(q^2)= \frac{f_+(0)}{1-\frac{q^2}{m_{\rm pole}^2}}
\label{eq:pole}
\eeq
is fitted.
Note that such an effective pole mass value has no clear
physical interpretation and that the proposed $q^2$ variation does not comply 
with constraints from QCD. The obtained value may nonetheless be useful for comparison
with results from different experiments.

\subsection{Quantitative expectations}
Values of the parameters that determine $f_+(q^2)$ were obtained initially
from constituent quark models and from QCD sum rules. These two approaches have
an intrinsically limited accuracy. In this respect, results from lattice QCD 
computations are more promising because their accuracy is mainly limited
by available computing resources.

\subsubsection{Quark Models}
Quark model calculations estimate meson wave functions and use them to
compute the matrix elements that appear in the hadronic current.
There are a large variety of theoretical calculations
\cite{ref:qm}.
Among these models we have selected the ISGW model \cite{ref:isgw}, 
simply because it is widely used
to simulate heavy hadron semileptonic decays.
This model was expected to be valid in the vicinity of $q^2_{\rm max}$,
a region of maximum overlap between  the initial
and final meson wave functions.
In ISGW2 \cite{ref:isgw2}
the exponential $q^2$ dependence of the form factor 
has been replaced by another
parameterization, with a dipole behavior, expected to be valid over a larger $q^2$ range: 
\beq
f_+^{{\rm ISGW2}}(q^2) = \frac{f_+(q^2_{\rm max})}{\left ( 1+\alpha_I(q^2_{\rm max}-q^2) \right )^2},~\alpha_I=\frac{1}{12}r^2.
\eeq
The predicted values of the parameters are
$f_+(q^2_{\rm max})=1.23$ and
$r=1.12~ \GeV^{-1}$ \cite{ref:isgw2}.

\subsubsection{QCD sum rules}
QCD sum rules \cite{ref:qcdsr} and their extension on the light cone
\cite{ref:lcqcdsr}, 
are expected to be valid at low 
$q^2$. 
Using a value
of 150 $\MeV$ for the strange quark mass, one obtains \cite{ref:lcqcdsr}:
\beq
f_+(0)=0.78\pm0.11 ~{\rm and}~ \alpha_{\rm pole}=-0.07^{+0.15}_{-0.07},
\eeq
using the modified pole ansatz.
The uncertainty of $f_+(0$) is estimated to be of order 15$\%$,  and the $q^2$ dependence 
is expected to be dominated by a single pole at the $D_s^*$ mass
because the value of $\alpha_{\rm pole}$ is compatible with zero.

\subsubsection{Lattice QCD}
Lattice QCD computation is the only approach able to compute
$f_+(q^2)$ from first principles. 
Current results must be
extrapolated to physical values of light quark masses and corrected
for finite lattice size and discritization effects.
There have been 
several evaluations of $f_+(q^2)$ for different values of the 
momentum transfer in the quenched approximation
\cite{ref:lqcd1,ref:lqcd2}. These results have been combined
\cite{ref:lqcd1}, giving $f_+(0)=0.73\pm0.07$. The first unquenched
calculation has been published recently \cite{ref:lqcd3}:
 $f_+(0)=0.73\pm0.03\pm0.07$ and $\alpha_{\rm pole}=0.50\pm0.04$,   
using the modified pole ansatz to parameterized the $q^2$
dependence of the form factor.

\subsection{Analyzed parameterizations}
The different parameterizations of $f_+(q^2)$ considered in this analysis 
are summarized in Table \ref{tab:expect}, along with
their corresponding parameters and expected values, where available.

{\small
\begin{table}
  \caption {Parameterizations of $f_+(q^2)$.}
\begin{center}
  \begin{tabular}{lcc}
    \hline\hline
modeling & parameters & expected values\\
\hline
z expansion\cite{ref:beforehill} & $a_0,~r_k=a_k/a_0$ & no prediction  \\
general 2-poles \cite{ref:bk}& $f_+(0),~\beta_{\rm pole},~\delta_{\rm pole}$   & no prediction \\
modified pole \cite{ref:bk}&$f_+(0),~\alpha_{\rm pole}$ & $\delta_{\rm pole} = 0$  \\
simple pole &$f_+(0),~m_{\rm pole}$  & $m_{\rm pole}=m_{D_s^*}$\\
ISGW2\cite{ref:isgw2} & $f_+(t_-),~\alpha_I$ & $f_+(t_-)=1.23$   \\
  &  & $\alpha_I=0.104~\GeV^{-2}$   \\
\hline\hline
  \end{tabular}
\end{center}
\label{tab:expect}
\end{table}
}

\section{The \babar\ Detector and Dataset}
\label{sec:babar}

A detailed description of the \babar\ detector and of the algorithms used
for charged and neutral particle reconstruction and identification is 
provided elsewhere~\cite{ref:babar, ref:babardet}. 
Charged particles are reconstructed by matching hits in 
the 5-layer double-sided silicon vertex tracker (SVT) 
with track elements in the 40-layer drift chamber (DCH), which is
filled with a gas mixture of helium and isobutane.
Slow particles which do not leave enough hits  
in the DCH due to the bending in the $1.5$-T
magnetic field, are reconstructed in the SVT.
Charged hadron identification is performed combining the measurements of 
the energy deposition in the SVT and in the DCH with the information from the
Cherenkov detector (DIRC). Photons are detected and measured in the 
CsI(Tl) electro-magnetic calorimeter (EMC). 
Electrons are identified by the ratio of the track momentum to the
associated energy deposited in the EMC, the transverse profile of the shower,
the energy loss in the DCH, and the Cherenkov angle in the DIRC.
Muons are identified in the instrumented flux return, composed
of resistive plate chambers interleaved with layers of steel and brass.

The results presented here are obtained using  
a total integrated luminosity of $75~\fb^{-1}$ 
registered by the \babar\ detector during the years 2000-2002. 
Monte Carlo (MC) simulation samples of $\Upsilon(4S)$ decays, charm and other light quark 
pairs from continuum equivalent, respectively, to  $2.8,~1.2~{\rm and}~0.7$ times the data statistics,
respectively, have 
been generated using  GEANT4 \cite{ref:geant4}. These are used mainly
to evaluate background components. Quark fragmentation, in continuum events, is described 
using the JETSET package \cite{ref:jetset}. 
The MC distributions have been rescaled to the 
data sample luminosity, using the expected cross sections 
of the different components ($1.3$nb for $c\overline{c}$, $0.525$
nb for $\Bp\Bm$ and $B^0 \bar{B}^0$, $2.09$nb for light $u\bar{u}$, $d\bar{d}$ and $s\bar{s}$ quark events).
Dedicated samples of pure
signal events, equivalent to seven times the data statistics,
are used to correct measurements for efficiency and 
finite resolution effects. They have been generated using the modified
pole parameterization ansatz for $f_+(q^2)$ with $\alpha_{\rm pole}=0.50$.
Radiative decays $(D^0 \rightarrow K^- e^+ \nu_e \gamma)$ are modeled by PHOTOS
\cite{ref:photos}. To account for one of the most important sources
of background,
a special event sample with, in each event, at least one cascade
decay $\Dstarp \rightarrow \Do \pi^+,~\Do \rightarrow \Km \pi^0 e^+ \nu_e$
(or its charge conjugate) has been generated with a parameterization
of the form factors in agreement with measurements from the FOCUS 
collaboration \cite{ref:focuskstar}. 
Events with a $\Dstarp$ and a $D^0$ decaying into $K^-\pi^+$ or
$K^-\pi^+\pi^0$ have been reconstructed in data and simulation.
These control samples have been used to adjust the
$c$-quark fragmentation distribution and the kinematic
characteristics of particles accompanying the $D$
 meson in order to better match the data.
They have been used also to measure the reconstruction accuracy on the missing
neutrino momentum.

\section{Signal reconstruction}
\label{sec:Analysis}

%
We reconstruct $D^0 \rightarrow K^- e^+ \nu_e (\gamma)$  decays in 
$e^+e^- \rightarrow c\bar{c}$  events where the $D^0$ originates from
the $D^{*+} \rightarrow D^0 \pi^+$.
The main sources of background arise from events with a kaon and 
electron candidate. Such events come from $\Upsilon(4S)$ decays and the continuum 
production of charmed hadrons. Their contribution is reduced
using variables sensitive to the particle production characteristics that are different
for signal and background events.

\subsection{Signal selection}
\label{sec:Analysisq2}
Charged and neutral particles are boosted to the center of mass system (c.m.) 
and the event thrust
axis is determined.
The direction of this axis is required to be in the interval
$|\cos(\theta_{{\rm thrust}})|<0.6$ to minimize the loss of particles in 
regions close to the beam axis.
A plane perpendicular to the thrust axis is used to define two hemispheres,
equivalent to the two jets produced by quark fragmentation. In each hemisphere, 
we search for pairs of oppositely charged leptons and kaons.
For the charged lepton candidates we consider 
only electrons or positrons with c.m. momentum
greater than 0.5 $\GeVc$.
 
Since the $\nu_e$ momentum is unmeasured, a kinematic fit is performed, constraining the invariant mass of the candidate
$e^+K^-\nu_e$ system to the $D^0$ mass. In this fit,
the $D^0$ momentum and the neutrino energy are estimated from the other particles 
measured in the event. The $D^0$ direction is taken as the direction opposite
to the sum of the momenta of all reconstructed particles in the event, except for the kaon and the
positron associated with the signal candidate. 
The energy of the jet is determined from the total c.m. energy and from the measured
masses of the two jets.
The 
neutrino energy is estimated as the difference between the total energy of the jet
and the sum of the energies of all reconstructed particles in the hemisphere. 
A correction, which depends on the value of the missing energy measured in the opposite jet,
is applied to account for the presence of missing energy due to 
particles escaping detection, even in the absence of a neutrino from the $D^0$ decay.

\begin{figure}[!htb]
\begin{center}
\includegraphics[height=11cm]{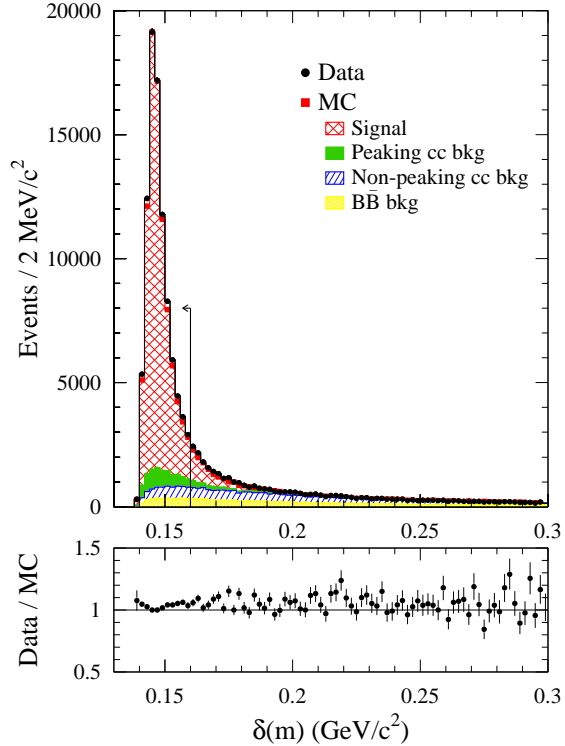}
\caption{ Comparison of the $\delta(m)$ distributions from data and simulated events. 
MC events have been normalized to the sample luminosity according to the 
different cross sections. An excess of background events of the order of 5$\%$
is observed for large values of $\delta(m)$. The arrow indicates the additional selection applied
for the $q^2$ distribution measurement.}
\label{fig:deltamback}
\end{center}
\end{figure}

The $D^0$ candidate is retained
if the $\chi^2$ probability of the kinematic fit exceeds 10$^{-3}$.
Detector performance for the reconstruction of the $\Do$ direction and for the missing
energy are measured using events in which the $\Do$ decays
into $\Km \pi^+$.
Corrections are applied to account for observed differences between data and 
simulation.
Each $D^0$ candidate is combined with a charged pion, with the  same charge as the lepton,
and situated in the same hemisphere.
The mass difference
$\delta(m) = m(\Do \pi^+)-m(\Do)$ is evaluated and is shown in Fig.\
\ref{fig:deltamback}. 
This distribution contains events which in addition pass the requirements on the Fisher discriminant 
$F_{B\bar{B}}$ suppressing
$B\bar{B}$ background and also give a satisfactory kinematic fit constraining the invariant mass .
This last requirement is the reason of the slow decrease of the
$\delta(m)$ distribution. At large $\delta(m)$ values, a small excess of background is measured in data and the simulation
is rescaled accordingly.
Only events with $\delta(m)<0.16~\GeVcd$ are used in the analysis.

\subsection{Background rejection}
\label{sec:backgrej}
Background events arise from $\Upsilon(4S)$ decays and hadronic 
events from the continuum. 
Three variables are used to reduce the contribution from $B\bar{B}$ events:
R$_2$ (the ratio between the second and zeroth
order Fox-Wolfram moments \cite{ref:r2}), the total charged and neutral 
multiplicity and the momentum of the soft pion ($\pi_s$) from the $\Dstarp$.
These variables exploit the topological differences between events with $B$ decays and 
events with $c\bar{c}$ fragmentation. The particle distribution in $\Upsilon(4S)$ decay events 
tends to be isotropic as the $B$ mesons are produced near threshold, while the distribution in $c\bar{c}$ events 
is jet-like as the c.m. energy is well above the charm threshold. This also results in a softer 
$D^{*+}$ momentum spectrum in $\Upsilon(4S)$ decays compared to $c\bar{c}$ events.

\begin{figure}[!htb]
  \begin{center}
    \mbox{\epsfig{file=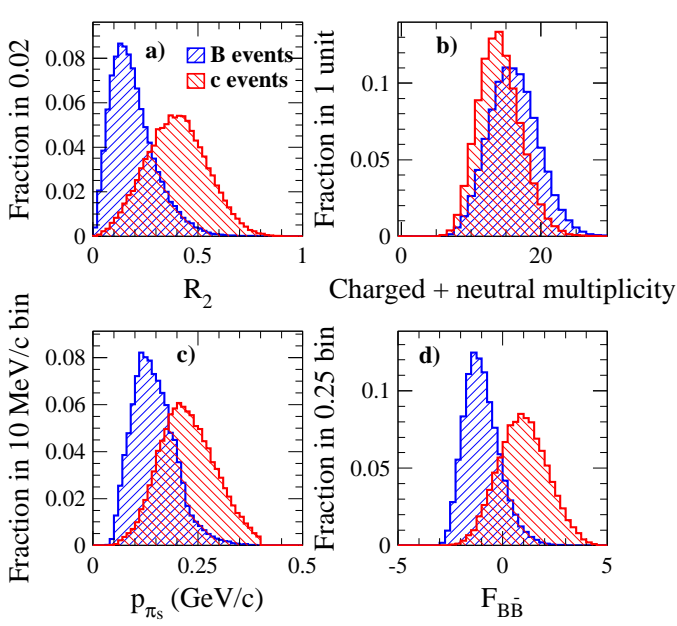,width=0.53\textwidth}}
  \end{center}
  \caption[]{ MC simulations of distributions of the variables used in the Fisher discriminant
analysis to reduce the $B\bar{B}$ event background: a) the normalized 
second Fox-Wolfram moment ($R_2$), b) the event 
particle multiplicity, c) the slow-pion momentum 
distribution, in the c.m. frame, d) the Fisher variable 
for $B\bar{B}$ and for charm signal events.}
   \label{fig:h2h0_mult}
\end{figure}

Corresponding distributions
of these variables for signal and background events are given in Fig.\ \ref{fig:h2h0_mult}. 
These variables have been combined linearly in a Fisher discriminant. 
The requirement $F_{B\bar{B}}>0.5$ retains $65\%$ of signal and 6$\%$ of $B\bar{B}$-background events.

Background events from the continuum arise mainly from charm particles
as requiring an electron and a kaon reduces the 
contribution from light-quark flavors
to a low level.
Because charm hadrons take a large
fraction of the charm quark energy charm decay products
have higher average energies and different angular distributions  (relative to 
the thrust axis or to the $D$ direction) compared with
other particles in the hemisphere emitted from the hadronization
of the $c$ and $\overline{c}$ quarks. These other particles are
referred to as ``spectator'' in the following; the 
``leading'' particle is the one with the largest momentum. 
To reduce background from $c \bar{c}$ events, the following variables are used:

\begin{figure}[!htb]
  \begin{center}
    \mbox{\epsfig{file=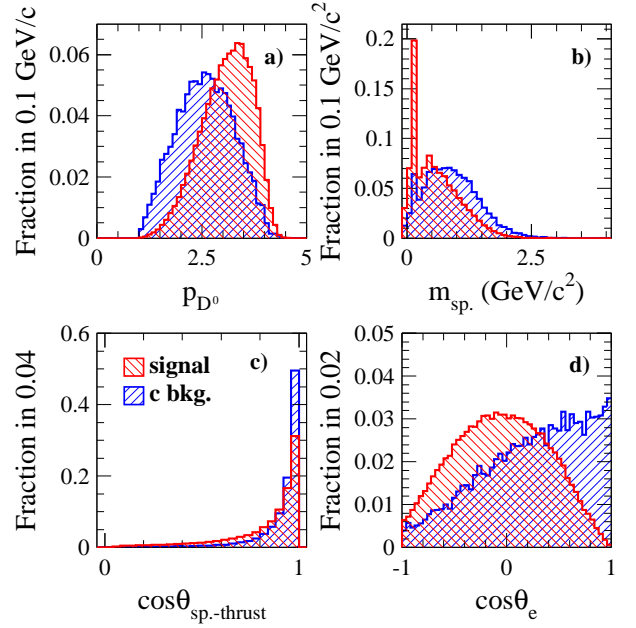,width=0.53\textwidth}}
  \end{center}
  \caption[]{ MC simulations of some of the variables used in the Fisher discriminant
analysis to reduce the $c\bar{c}$-event background: a) the $D^0$ momentum after the kinematic fit,
b) the mass of the spectator system (peaks, at low mass values correspond
to events with a single charged pion or photon reconstructed in the
spectator system), c) the cosine of the angle between the spectator system momentum
and the thrust direction, d) the cosine of the angle of the positron direction, relative to the kaon
direction, in the $e\nu_e$ c.m. frame.
}
   \label{fig:fisher1}
\end{figure}

\begin{figure}[!htb]
  \begin{center}
    \mbox{\epsfig{file=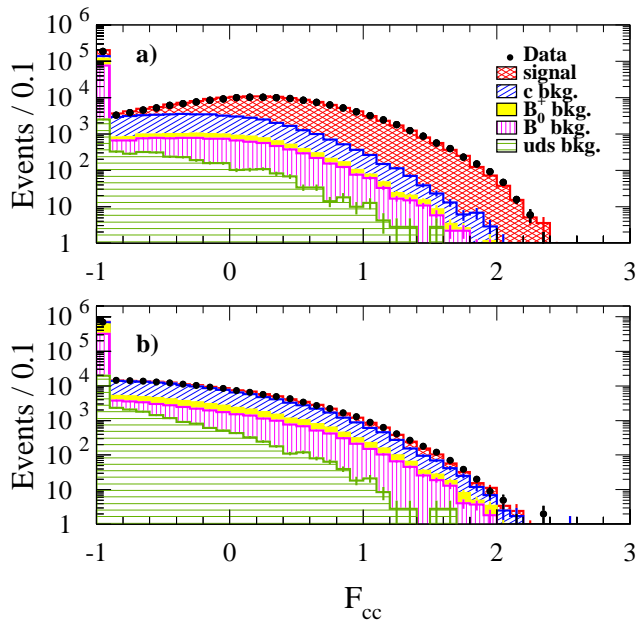,width=.54\textwidth,angle=0}}
  \end{center}
  \caption[]{Distribution of the values of the Fisher variable
in the signal region ($\delta(m)<0.16~\GeVcd$ in a), and for masses above
the signal region ($\delta(m)>0.16~\GeVcd$ in b).}
  \label{fig:fisherkenu}
\end{figure}

\begin{itemize}
\item the $\Do$ momentum;
\item the spectator system mass, $m_{{\rm sp.}}$, 
which has lower values for signal events;
\item the direction of the spectator system momentum relative to the thrust axis $\cos{\theta_{{\rm sp.-thrust}}}$;
\item the momentum of the leading spectator track;
\item the direction of the leading spectator track relative to the $\Do$ direction;
\item the direction of the leading spectator track relative to the thrust axis;
\item the direction of the lepton relative to the kaon direction, in the 
dilepton rest frame, $\cos{\theta_e}$;
\item the charged lepton momentum, $p_e$, in the c.m. frame.
\end{itemize}
The first six variables depend on the properties of
$c$-quark hadronization whereas the last two are related to decay
characteristics of the signal. Distributions for four of the
most discriminating variables
are given in Fig.\ \ref{fig:fisher1}.
$\Do \rightarrow K^- \pi^+$ events have been used to tune the
simulation parameters so that distributions of the variables used to reject 
background
agree with those measured with data events.
These eight variables have been combined linearly into a Fisher discriminant 
variable ($F_{cc}$)
and events have been kept for values above 0.
This selection retains $77\%$ of signal events that were kept by the previous selection requirement and
rejecting $66\%$ of the background (Fig.\ \ref{fig:fisherkenu}).

The remaining background from $c\bar{c}$-events can be divided into peaking (60$\%$) and 
non-peaking (40$\%$) candidates. 
Peaking events are those background events whose distribution is peaked around the signal region.
These are mainly
events with a real $\Dstarp$ in which the slow $\pi^+$ is
included in the candidate track combination. 
Backgrounds from $e^+e^-$ annihilations into light $u\bar{u},~d\bar{d},~s\bar{s}$ quarks and $B\bar{B}$ events
are non-peaking.
These components, from simulation, are displayed
in Fig.\ \ref{fig:deltamback}.

\subsection{\boldmath{$q^2$} measurement}

\begin{figure}[!htb]
\begin{center}
\includegraphics[height=10cm]{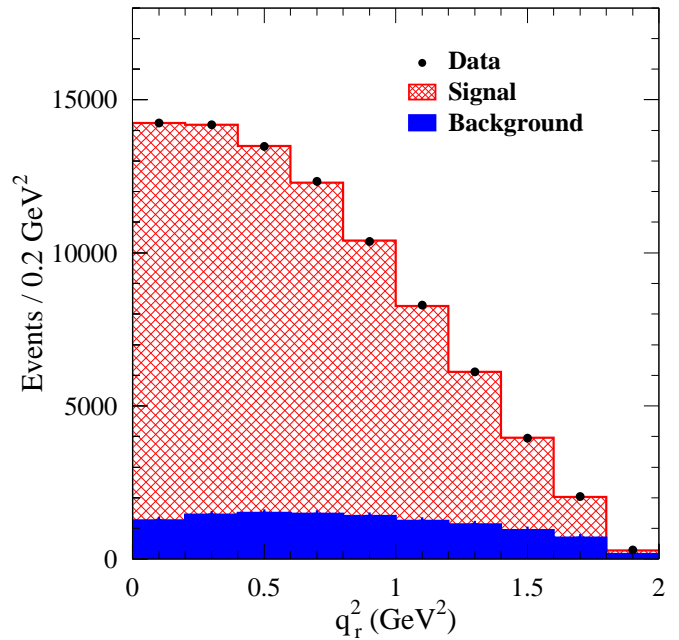}
\caption{ The measured $q^2_r$ distribution (data points) compared to the sum of 
the estimated background and of the fitted signal components.
}
\label{fig:q2rall}
\end{center}
\end{figure}

To improve the accuracy of the reconstructed $D^0$ momentum, 
the nominal $\Dstarp$ mass 
is added as a constraint in the previous fit and only events with
a $\chi^2$ probability higher than 1$\%$ are kept 
(Fig.\ \ref{fig:deltamback} is obtained requiring only that the fit
has converged). 
The measured $q^2_r$ distribution, 
where $q^2_r= \left ( p_D-p_K \right )^2$,
is given in Fig.\ \ref{fig:q2rall}.
There are 85260 selected $D^0$ candidates containing an estimated number
of 11280 background events.
The non-peaking component comprises $54\%$ of the background.

To obtain the true $q^2$ distribution, the measured one
has to be corrected for selection efficiency
and detector resolution effects. This is done using an unfolding algorithm
based on MC simulation of these effects.  

\begin{figure}[!htb]
\begin{center}
\includegraphics[height=9cm]{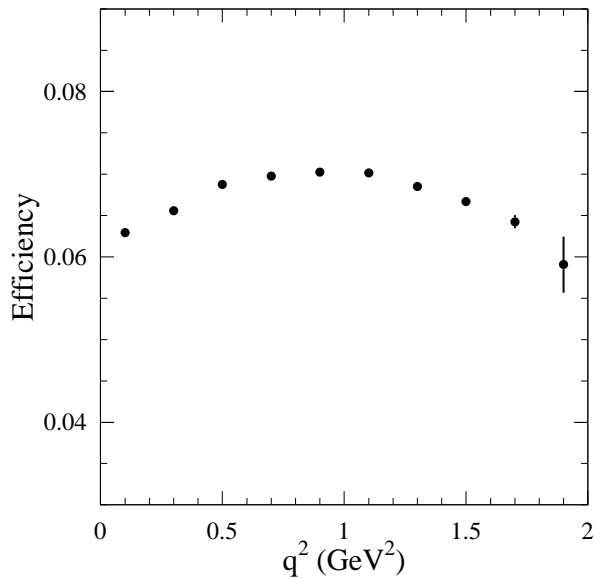}
\caption{ The efficiency as a function of $q^2$, measured with
simulated signal events, after all selection criteria 
applied.}
\label{fig:effi}
\end{center}
\end{figure}

The variation of the selection efficiency as a function of $q^2$
is given in Fig.\ \ref{fig:effi}.
The resolution of the $q^2$ measurement for signal events is obtained 
from MC simulation. The resolution function can fitted  
by the sum of two Gaussian functions, 
 with standard deviations 
$\sigma_1=0.066 ~\GeV^2$ and $\sigma_2=0.219 ~\GeV^2$, respectively.
The narrow component corresponds to 40$\%$ of the events.

To obtain the unfolded $q^2$ distribution for signal events, 
corrected for resolution 
and acceptance effects,
the Singular Value Decomposition (SVD) 
\cite{ref:svd} of the resolution matrix
has been used.
This method uses a
two-dimensional matrix which relates the generated $q^2$ distribution to the detected 
distribution, $q_r^2$, as input.
After subtracting the estimated background contribution, the measured binned $q^2_r$ distribution
is linearly transformed into the unfolded 
$q^2$ distribution. 
This approach provides the full covariance matrix for the
bin contents of the unfolded distribution.
Singular values (SV) are ordered by decreasing values. 
These values contain the information needed to transform the measured
distribution into the unfolded spectrum, along with statistical uncertainties
from fluctuations. 
Not all SV are relevant; non-significant values have 
zero mean and standard deviation equal to unity \cite{ref:svd}.
Using toy simulations, we find
that seven SV have to be kept
with events distributed over ten bins.
Because the measurement of  the form-factor parameters relies on the measured $q_r^2$  distribution, 
it does not require unfolding, and is independent 
of this particular choice.

\section{Results on the \boldmath{$q^2$} dependence of the hadronic form factor }
\label{sec:Physics}

\begin{figure}[!htb]
\begin{center}
\includegraphics[height=9cm]{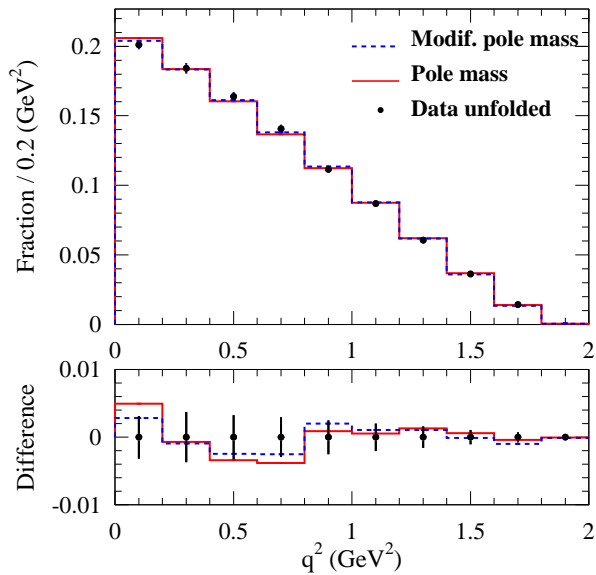}
\caption{ Comparison between the normalized unfolded $q^2$ distribution 
obtained from this analysis,
and those
corresponding to two fitted models. Lower plot gives the difference
between measured and fitted distributions.
The error bars represent statistical errors only.
}
\label{fig:q2datasim}
\end{center}
\end{figure}

The unfolded $q^2$ distribution, normalized to unity, is presented in Fig.\ \ref{fig:q2datasim}
and in Table \ref{tab:errmeas}.  
Also given in this table are the statistical and total uncertainties and the correlations of the data in the ten bins.  
Figure \ref{fig:q2datasim} shows the result of fits to the data for two parameterizations of the form factor 
with a single free parameter, the simple pole and the modified pole ansatz. 
Both fitted distributions agree well with the data.  

A summary of these and other form factor parameterizations is given in Table \ref{tab:fittedparam}.   
These results will be discussed in detail in Section \ref{sec:comper}.

\begin{table*}[htbp]
  \caption {Statistical and total uncertainty matrices for the
normalized decay distribution (corrected for 
acceptance and finite resolution effects) in ten 
bins of $q^2$ from $0$ to $2~\GeV^2$, and for the 
ratio $R_D$ (see Section \ref{sec:rate}).
The total decay distribution
has been normalized to unity for $q^2$ varying over ten $0.2~\GeV^2$ intervals.
The uncertainty matrices are provided for both the statistical (upper half) and total (lower half) uncertainties. 
The uncertainty
on each measured value is given along the diagonal. Off-diagonal terms correspond to the 
correlation coefficients. 
}
\begin{center}
\small{
  \begin{tabular}{lccccccccccc}
    \hline\hline
$q^2$ bin $(\GeV^2)$& &[0, 0.2] &[0.2, 0.4] &[0.4, 0.6] &[0.6, 0.8] &[0.8, 1.0] &[1.0, 1.2] &[1.2, 1.4] &[1.4, 1.6] &[1.6, 1.8] &[1.8, 2.0] \\
\hline
$R_D$ and&0.9269 &  &      &    &      &  & &  & &  & \\
 fractions& &0.2008   &   0.1840   &    0.1632   &    0.1402    &   0.1122 &
 0.0874  & 0.0602  & 0.0367  & 0.0146  & 0.0007\\
\hline
statistical &0.0072 &0.166  & 0.122 & 0.111& 0.107 & 0.107 & 0.102 & 0.101 &0.116 & 0.081 &0.060 \\
uncertainties & &0.0031 & -0.451 & -0.155& 0.117 & 0.005 & 0.023 & 0.002 &0.002 & -0.002 &-0.002 \\
and & & & 0.0037 & -0.225 & -0.304 & 0.095 & 0.041 & -0.025 & -0.005 & 0.005 &-0.005 \\
correlations & && & 0.0033 & -0.155 & -0.345 & 0.079 & 0.058 & -0.018 & -0.010 & -0.006 \\
 & & & & & 0.0029 & -0.113 & 0.352 & 0.058 & 0.066 & -0.013 & -0.024 \\
& & & & & & 0.0025 & 0.073 & -0.345 & 0.018 & 0.075 & 0.067 \\
& & & & & & & 0.0020 & -0.029 & -0.329 &-0.004 &0.060 \\
& & & & & & & & 0.0016 & 0.110 & -0.339 & -0.347 \\
& & & & & & & & & 0.0011 & 0.217 & 0.012 \\
& & & & & & & & & & 0.00075 & 0.965 \\
& & & & & & & & & & & 0.000057 \\
    \hline
\hline
total &0.0139 &-0.154 & 0.072 & 0.034 & 0.093 & 0.046 & 0.084 & 0.231 &0.278 & 0.210 &0.184 \\
uncertainties & &0.0041 & -0.462 & -0.257 & 0.022 & 0.099 & 0.089 &-0.232 &-0.092 & -0.056 &-0.048 \\
and & && 0.0040 & -0.102 & -0.247 & 0.005 & 0.020 & 0.015 & -0.035 & -0.055 & -0.048 \\
correlations & && & 0.0035 & -0.122 & -0.377 & 0.050 & 0.074 & -0.043 & -0.033 & -0.027 \\
&& & & & 0.0030 & -0.127 & 0.320 & 0.092 & 0.056 & -0.038 & -0.048 \\
& & & & & & 0.0026 & 0.041 &-0.331 & 0.033 & 0.089 & 0.079 \\
& & & & & & & 0.0022 & 0.084 & -0.264 &0.010 &0.057 \\
& & & & & & & & 0.0018 & 0.159 & -0.235 & -0.254 \\
& & & & & & & & & 0.0012 & 0.369 & 0.194 \\
& & & & & & & & & & 0.0009 & 0.973 \\
& & & & & & & & & & & 0.000065 \\
    \hline\hline
  \end{tabular}}
\end{center}
  \label{tab:errmeas}
\end{table*}

\begin{figure}[!htb]
  \begin{center}
    \mbox{\epsfig{file=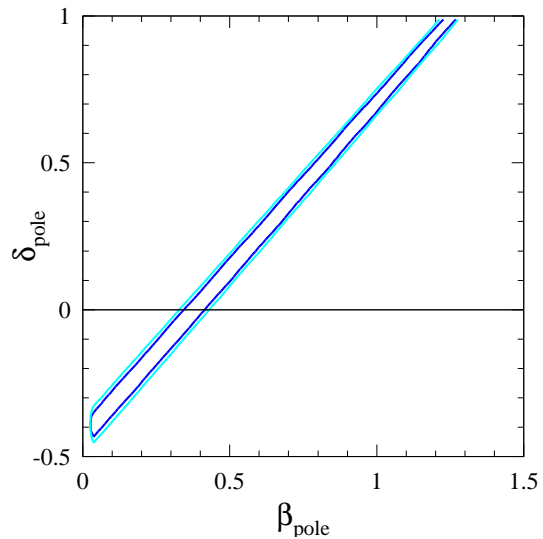,width=8cm,height=8cm}}
  \end{center}
  \caption[]{Contours at 70$\%$ and 90$\%$ CL resulting from
the fit of the parameterization of the form factor $q^2$ dependence
with two parameters as given in Eq. (\ref{eq:twopoles}). The value
$\delta_{\rm pole}=0$ corresponds to the modified pole ansatz.}
  \label{fig:2dfit}
\end{figure}

The fit to a model is done by
comparing the number of events measured in a given bin of $q^2$
with the expectation from the exact analytic integration
of the expression $\left | \vec{p}_K (q^2) \right |^3 \left |f_+(q^2) \right |^2$
over the bin range, with the overall normalization left free. 
The result of the fit corresponding to the parameterization of the form
factor using two parameters (see Eq. \ref{eq:twopoles}) is given in Fig.\ \ref{fig:2dfit}. 
\begin{table*}
  \caption {Fitted values of the parameters corresponding to 
different parameterizations of $f_+(q^2)$. The last column
gives the $\chi^2/NDF$ of the fit when using the value expected for the
parameter.}
\begin{center}
  \begin{tabular}{lclcl}
    \hline\hline
Theoretical &Unit & Parameters&$\chi^2$/NDF& Expectations \\
ansatz & & & & [$\chi^2$/NDF]\\
\hline
$z$ expansion& & $r_1  = -2.5 \pm 0.2 \pm 0.2$ &5.9/7 &  \\
   &                    & $r_2  = 0.6 \pm 6. \pm 5.$& &\\

Modified pole &&$\alpha_{\rm pole} = 0.377 \pm 0.023 \pm 0.029$ & 6.0/8& \\

Simple pole &$\GeVcd $ &$m_{\rm pole} = 1.884 \pm 0.012 \pm 0.015$&7.4/8& $2.112~[243/9]$\\

ISGW2 & $\GeV^{-2} $& $\alpha_I = 0.226 \pm 0.005 \pm 0.006$ &6.4/8 &$0.104~[800/9]$\\
\hline \hline
  \end{tabular}
\end{center}
  \label{tab:fittedparam}
\end{table*}

\subsection{Systematic Uncertainties}
\label{sec:Systematics}

Systematic uncertainties of the form factor parameters are likely to originate from 
imperfect simulation of $c$-quark fragmentation and the detector response, from 
uncertainties in the background composition and the individual contributions for the 
selected signal sample, the uncertainty in the modeling of the signal decay and the 
measurement of the $q^2$ distribution.
We study the origin and size of various systematic 
effects, correct the MC simulation, if possible, assess the impact of the 
uncertainty of the size of correction on the fit results, and adopt 
the observed change as a contribution to the systematic uncertainty on the 
fitted parameters for the different parameterizations under study. 
Some of these studies make use of standard \babar\ evaluations of detection efficiencies, 
others rely on special data control samples, for instance hadronic decays 
$D^0 \rightarrow K^-\pi^+$ or $K^- \pi^+ \pi^0$.

\subsubsection{$c$-quark hadronization tuning}

The signal selection is based on variables related to $c$-quark
fragmentation and decay properties  of signal events.
Simulated events have been weighted 
to agree with the distributions observed in data.
Weights have been obtained using events with a reconstructed
$\Do$ decaying into $\Km \pi^+$. 
After applying these corrections, the distribution
of the Fisher discriminant that contains
these variables is compared for data and simulation.
The remaining difference is used
to evaluate the  corresponding systematic uncertainty. It corresponds
to the variations on fitted quantities obtained
by correcting or not for this difference, which is
below 5$\%$ over the range of this variable. 

\subsubsection{Reconstruction algorithm}
It is important to verify that the $q^2$ variation of the selection 
efficiency is well described by the simulation. 
This is done by analyzing $\Do \rightarrow \Km \pi^+ \pi^0$
as if they were $\Km e^+ \nu_e$ events.
The two photons from the $\pi^0$ are removed and 
events are reconstructed using the algorithm applied to the semileptonic
$\Do$ decay. The ``missing'' $\pi^0$ and the charged pion 
play, respectively, the roles of the neutrino and  the electron. 
To preserve the correct kinematic
limits, it is necessary to take into account that the ``fake'' neutrino has the
$\pi^0$ mass and that the ``fake'' electron has the $\pi^+$ mass.

Data and simulated events, 
which satisfy the same analysis selection criteria as for $K e \nu_e$, have been compared.
For this test, the $\cos(\theta_e)$ and $p_e$ are removed from the Fisher discriminant, 
because distributions for these two variables are different from the signal events. 

The ratio of efficiencies measured in data and simulation is 
fit with a linear expression in $q^2$. The corresponding slope,
$(0.71 \pm 0.68)\%$, indicates that there is no
significant bias when the event selection criteria are applied. 
The measured slope is used to define a correction and to estimate the corresponding systematic uncertainty.

\subsubsection{Resolution on $q^2$}

\begin{figure}[!htb]
  \begin{center}
     \epsfig{file=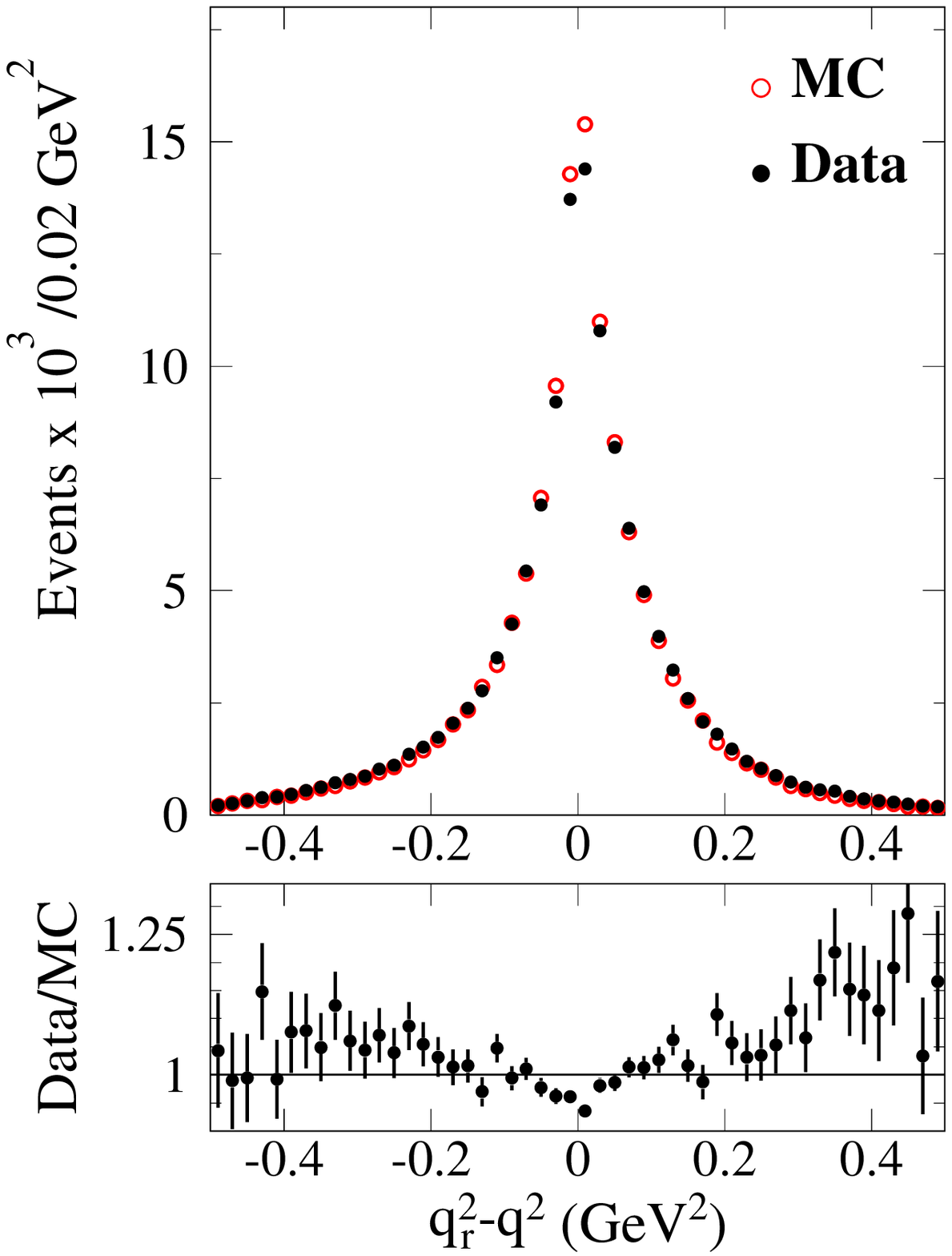,height=9.5cm} \\
     \epsfig{file=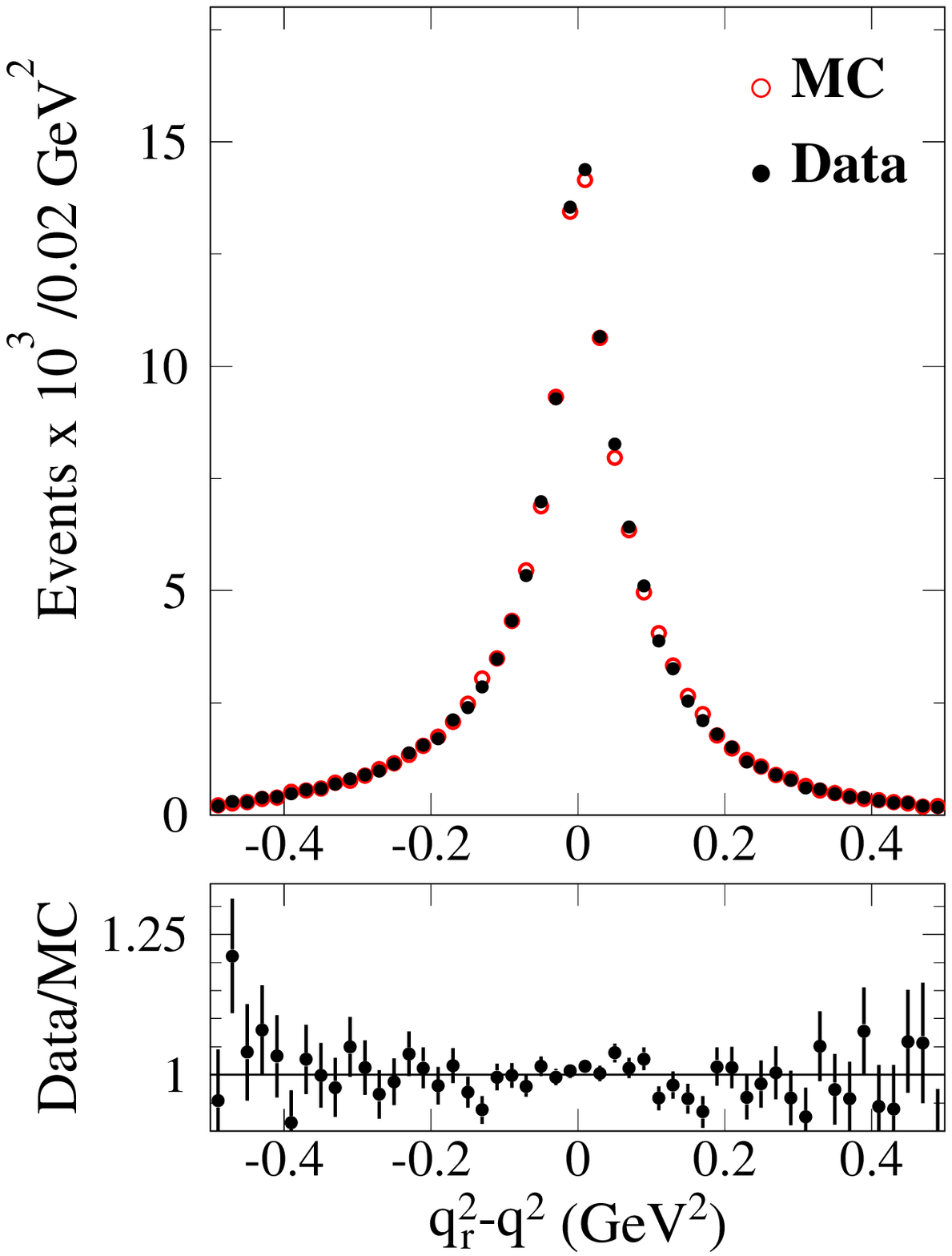,height=9.5cm}
  \end{center}
  \caption[]{ Distribution of the difference between the true and the
reconstructed value of $q^2$. 
$D^0 \rightarrow K^- \pi^+ \pi^0$ data events correspond to dark squares and open circles are used for simulated ones. 
Ratio (Data/MC) of the two distributions are displayed. 
The distributions in a) compare data and simulated
events before applying corrections measured with
$D^0 \rightarrow K^-\pi^+$  events, whereas these corrections have been applied for plots in b).
}
   \label{fig:resol_q2_kpipi0}
\end{figure}

To measure possible differences between data and simulation
on the $q^2$ reconstruction accuracy,
$\Do \rightarrow \Km \pi^+ \pi^0$ events are used again. 
Distributions 
of the difference $q^2_r-q^2$, obtained by selecting events in a given
bin of $q^2$ are compared. These distributions
are systematically slightly narrower for simulated events
and the fraction of events in the distant tails are higher for data
(see Fig.\ \ref{fig:resol_q2_kpipi0}).

With the $\Do \rightarrow \Km \pi^+$ sample we study,
in data and simulation, the
accuracy of the $\Do$ direction and missing energy reconstruction
for the $\Do \rightarrow K^- e^+ \nu_e$ analysis.
This information is used in
the mass-constrained fits and thus influences the $q^2$ reconstruction.
Once the simulation is tuned to reproduce the results obtained on data
for these parameters, the $q^2$ resolution distributions agree
very well, as shown in Fig.\ \ref{fig:resol_q2_kpipi0}.
One half of the measured variation on the fitted parameters from these corrections 
has been taken as a systematic uncertainty.

\subsubsection{Particle identification}
Effects from a momentum-dependent difference between data
and simulated events on
the charged lepton and on the kaon
identification have been evaluated.
Such differences, which are typically below 2$\%$, have been 
measured for selected, high purity samples of electrons and kaons.
These corrections have been applied and the observed variation
has been taken as the estimate of the systematic uncertainty.

\subsubsection{Background estimate}
\label{sec:backgkenu}

The background under the $\Dstarp$ signal has two components that have,
respectively, non-peaking and peaking behavior.

The non-peaking background originates from non-$c\overline{c}$ events
and from continuum charm events in which the $\pi_s$ candidate does not
come from a cascading $\Dstarp$. By comparing 
data and simulated event rates for $\delta(m)>0.18~\GeVcd$
(see Fig.\ \ref{fig:deltamback}),
a correction of 1.05 is determined from simulation
for the non-peaking background. This
correction is applied  and an
uncertainty of
$\pm 0.05$ is used as the corresponding systematic uncertainty.

Events which include a slow 
pion originating from $\Dstarp$ decay
contribute in several ways to the peaking background.
The production rate of $\Dstarp$ mesons in the
simulation is in agreement with expectations 
based on measurements from CLEO \cite{ref:cleocc}.
The uncertainty of $\pm 0.06$ on this comparison is dominated by the 
systematic uncertainty from the CLEO result.

To study the remaining effects, the peaking background components have been 
divided according to the 
process from which they originate 
and have been ordered by decreasing level of importance:

\begin{itemize}
\item {\it the $\Km$ and the electron originate from a $\Do$ decay ($54\%$).}
The main source comes from $\Do \rightarrow \Km \pi^0 e^+ \nu_e$.
We have corrected the decay branching fraction used for this channel
in the MC ($2.02 \%$) using recent measurements
($2.17\pm0.16\%$ \cite{ref:pdg06}). The uncertainty on this value
has been used to evaluate the corresponding systematic uncertainty. 

\item {\it the electron comes from a converted photon
or a Dalitz decay ($24\%$).} It has been assumed that the simulation 
correctly accounts for this component;

\item {\it the $K^-$ does not originate from a charm 
hadron ($14\%$).} This happens usually
when there is another negative charged kaon accompanying the $\Dstarp$. We 
have studied the production of charged kaon  accompanying a $\Dstarp$
using $\Do \rightarrow \Km \pi^+$ events  and measure a correction
factor of $0.87\pm0.02$ and $0.53\pm0.02$, respectively, for same sign
and opposite sign $K$-$D^*$ pairs. The 
simulation is modified accordingly and the remaining systematic uncertainty 
from this source becomes negligible;

\item {\it fake kaon candidate (mainly pions) ($6\%$) or fake electrons ($1\%$).}
Differences between data and MC on the evaluation of fake rates have been
studied in \babar.  As this affects small components of the total peaking
background rate, the effect of these differences has been neglected.
\end{itemize}

\subsubsection{Fitting procedure and radiative events}

To fit form factor parameters we compare the number of expected events
in each bin with the measured one after all corrections.
In this approach it is always assumed 
that the $q^2$ variation of $f_+(q^2)$ is given exactly
by the form factor parameterization.
This hypothesis is not correct, a priori, for radiative decays as 
$q^2=(p_D-p_K)^2 =(p_e+p_{\nu}+p_{\gamma})^2$ is not (perhaps)
equal to the variable that enters in  $f_+$ for such decays.
PHOTOS is used to generate decays with additional photons and
the modified pole ansatz is taken to parameterize the
hadronic form factor in signal events.
To quantify possible distortion of the fit
we compare the fitted value of a form factor parameter
with the one obtained from a fit to the generated $q^2$ distribution (see Table \ref{tab:corr}).

\begin{table}[!htb]
  \caption[]{ {Measured differences between the nominal and fitted values of  
parameters. Quoted uncertainties correspond to MC statistics. The last column 
gives the impact of the radiative effects on the form factor measurements as predicted by PHOTOS.} 
\label{tab:corr}}
{\small
\begin{center}
  \begin{tabular}{lcc}
    \hline\hline
Parameter & Measured difference& Bias from \\
    & (true-fitted)& radiation\\
\hline
$\delta(r_1)~(\times0.01)$& $1.2\pm13.0$ & $-0.4\pm 2.7$\\
$\delta(r_2)$ &  $+1.7\pm3.6$&  $+1.9\pm0.7$\\
$\delta(\alpha_{\rm pole})~(\times0.01)$&  $-1.2\pm1.4$& $-1.1\pm 0.3$\\
$\delta(m_{\rm pole})~( \MeVcd) $ & $+4.5\pm 6.3$ & $+4.6\pm 1.4$\\
$\delta(\alpha_I)~ ( \times 0.001\GeV^{-2}) $& $-2.7\pm3.1$ & $-2.4\pm 0.7$\\
    \hline\hline
  \end{tabular}
\end{center}
}
\end{table}
Corresponding corrections, given in the second column
of Table \ref{tab:corr}, have been applied and quoted uncertainties
enter in the systematic uncertainty evaluation.

To evaluate the importance of corrections induced by radiative effects,
we have compared also the fitted value of a parameter on $q^2$ distributions
generated with and without using PHOTOS. Measured differences are given in the
last column of Table \ref{tab:corr}. They have not been applied to
the values quoted in Table \ref{tab:fittedparam} for the different parameters.
We measure also that radiative effects affect mainly the fraction of the decay
spectrum in the first bin in Table \ref{tab:errmeas} which has to be increased by 
0.0012 to correct for this effect.

\begin{table*}[!htb]
  \caption[]{ {Summary of systematic uncertainties on the fitted 
parameters.}  \label{tab:systall}}
{\small
\begin{center}
  \begin{tabular}{lccccc}
    \hline\hline
Source & $\delta \left ( m_{\rm pole}\right )$ & $\delta \alpha_{\rm pole}$& $\delta \left ( \alpha_I\right )$& $\delta \left ( r_1\right )$ & $\delta \left (r_2\right )$ \\
       &  $(\MeVcd)$ & $(\times 0.01)$& $(\times 0.001~\GeV^{-2})$& $(\times 0.01)$&\\
\hline
$c$-hadronization tuning & $3.0$ & $ 0.6$ & $ 1.1$ & $ ~6.7$ & $ 1.4$   \\
Reconstruction algorithm & $ 7.8$ & $ 1.6$ & $ 3.1$ & $ ~6.5$ & $ 0.2$  \\
Resolution on $q^2$  & $ 3.4$ & $ 0.7$& $ 1.4$ & $ ~3.0$ & $ 2.1$   \\
Particle ID  & $ 5.5$ & $ 1.1$ & $ 2.3$ & $ ~3.8$ & $ 0.4$  \\
Background estimate  & $  7.7$ & $  1.4$& $ 3.0$ & $ 12.7$ & $ 2.0$   \\
Fitting procedure &  $6.3$ & $1.4$ & $3.1 $ & $13.0 $ & $3.6 $  \\
    \hline
Total  & $ 15.0$    & $  2.9$ & $ 6.0$ & $ 21.0$ & $ 4.9$ \\
    \hline\hline
  \end{tabular}
\end{center}
}
\end{table*}

\subsubsection{Control of the statistical accuracy in the 
SVD approach}
Once the number of SV is fixed, one must verify that the statistical 
precision obtained for each binned unfolded value is correct and if 
biases generated by removing information  are under control.
These studies are done with toy simulations.
One observes that the uncertainty obtained from a fit of the unfolded
distribution is underestimated by a factor which depends on the statistics of 
simulated events and is $\sim 1.06$ in the present analysis.
Pull distributions indicate also that the unfolded values, in each bin, 
have biases which are below 10$\%$ of the statistical uncertainty.
Similar studies are done for the determination of form factor parameters.

\subsubsection{Summary of systematic errors}
The systematic uncertainties for 
determining form factor parameters are summarized
in Table \ref{tab:systall}.

The systematic error matrix for the ten unfolded values is
computed by considering, in turn, each source of uncertainty and by measuring 
the
variation, $\delta_i$, of the corresponding unfolded value in each bin
($i$). The elements of the uncertainty matrix are the sum, over all sources of
systematic uncertainty, of the quantities $\delta_i \cdot \delta_j$. 
The total error matrix is evaluated as the sum of the matrices
corresponding, respectively, to statistical and systematic uncertainties.

\subsection{Comparison with expectations and with other measurements}
\label{sec:comper}
The summary of the fits to the normalized $q^2$ distributions are presented in Table \ref{tab:fittedparam}.
As long as we allow the form factor parameters to be free in the fit, the fitted 
distributions agree well with the data and it is not possible to reject any of the parameterizations.

However, if the form factor parameters are constrained to specific predicted values, the agreement 
is not good. For the ISGW2 model, the predicted  dependence of the form factor on $q^2$ disagrees 
with the data and the fitted value of the  parameter $\alpha_I$ differs from the predicted value, 
$\alpha_I =0.104~\GeV^{-2}$ by more than a factor two.

As observed by previous experiments, the simple pole model
ansatz, with $m_{\rm pole}=m_{D_s^*}=2.112~\GeVcd$ does not reproduce
the measurements. This means that the contribution from the continuum
$DK$ interaction cannot be neglected. 
If one introduces a second parameter $\delta_{{\rm pole}}$ to account for contributions from an effective 
pole at higher mass (see Eq. \ref{eq:twopoles}) the two parameters are fully correlated and there is 
no unique solution, as illustrated in Fig.\ \ref{fig:2dfit}. 
The modified pole ansatz corresponds to $\delta_{{\rm pole}}=0$.

In Table \ref{tab:results}  the fitted parameters for the simple pole ansatz and the modified pole 
\cite{ref:bk} ansatz are 
compared for different experiments.   The fitted pole masses are all well below the mass of the $D^*_s$ meson.  
The results presented here are consistent within the stated uncertainties with earlier measurements.
Except for the BELLE measurement,
all other measurements appear to favor a value of $\alpha_{{\rm pole}}$ that is lower than the value predicted 
by lattice QCD, namely $\alpha_{{\rm pole}}= 0.50\pm 0.04$.

\begin{table}[!htb]
  \caption {Fitted values for the parameters corresponding respectively
to a pole mass and a modified pole mass model for the form factor.
}
\begin{center}
{\small
  \begin{tabular}{lcc}
    \hline\hline
Experiment & $m_{\rm pole}$ $(\GeVcd)$ & $\alpha_{\rm pole}$   \\
\hline
CLEO \cite{ref:cleo3}  & $1.89 \pm0.05^{+0.04}_{-0.03}$ & $0.36 \pm0.10^{+0.03}_{-0.07}$  \\

FOCUS  \cite{ref:focus} & $1.93 \pm0.05 \pm 0.03$ & $0.28 \pm0.08 \pm0.07 $  \\

BELLE   \cite{ref:belle}& $1.82 \pm0.04 \pm 0.03$ & $ 0.52\pm0.08 \pm0.06 $  \\

\hline
This analysis  & $1.884 \pm 0.012 \pm 0.015 $ & $ 0.38\pm0.02 \pm0.03 $   \\
    \hline \hline
  \end{tabular}
}
\end{center}
\label{tab:results}
\end{table}

\begin{figure}[!htb]
\begin{center}
\includegraphics[height=9cm]{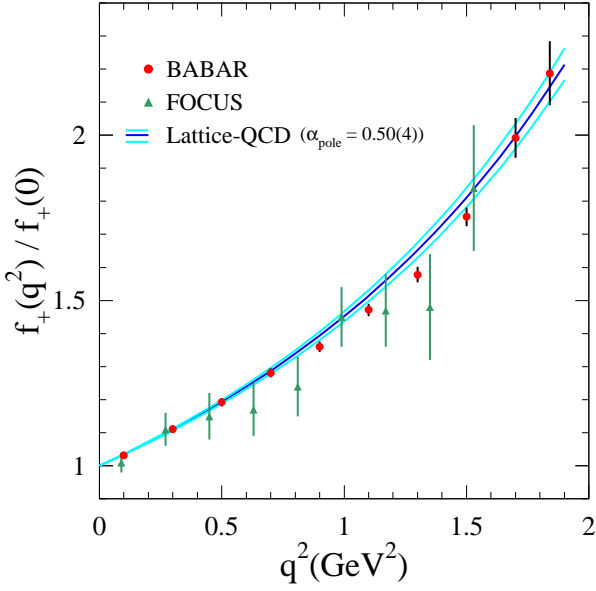}
\caption{ Comparison of the measured variation of $f_+(q^2)/f_+(0)$
obtained in the present analysis and in the FOCUS experiment \cite{ref:focus}.
The band corresponds
to lattice QCD \cite{ref:lqcd3} with the estimated uncertainty.}
\label{fig:fq2}
\end{center}
\end{figure}

In Fig.\ \ref{fig:fq2}, the dependence of the form factor on $q^2$ is presented.  The data are compared 
to earlier measurements by the FOCUS experiment, as well as with predictions from lattice 
QCD calculations \cite{ref:lqcd3}.  As stated above, the data favor a somewhat lower value for $\alpha_{{\rm pole}}$.

\begin{figure}[!htb]
\begin{center}
\includegraphics[height=9cm]{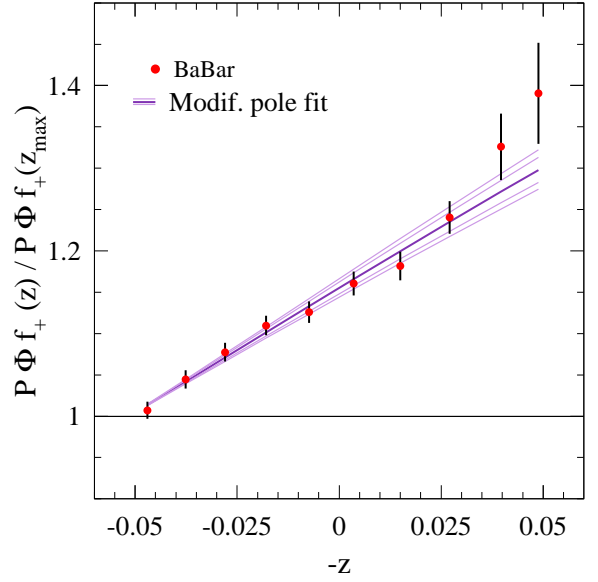}
\caption{ Measured values
for $P \times \Phi \times f_+$ are plotted versus $-z$ and 
requiring that $P \times \Phi \times f_+ = 1$ for $z=z_{\rm max}$.
The straight lines represent the result for the modified pole ansatz, 
the fit in the center and the statistical and total uncertainty.}
\label{fig:zhill}
\end{center}
\end{figure}

The data have also been mapped into the variable $z$.
Figure \ref{fig:zhill} shows the product $P \times \Phi \times f_+$ 
as a function of $z$.  By convention, this quantity 
is constrained to unity at $z=z_{\rm max}$, which corresponds to $q^2=0$.
We perform a fit to a polynomial, $P \times \Phi \times f_+ \sim 1+ r_1z + r_2z^2$.
The data are compatible with a linear dependence, which is fully consistent with the modified 
pole ansatz for $f_+(q^2)$, as illustrated in Fig.\ \ref{fig:zhill}.

\section{Branching fraction measurement}
\label{sec:rate}
The $\Do \rightarrow K^- e^+ \nu_e$ branching fraction
is measured relative to the reference decay channel,
$D^0 \rightarrow K^- \pi^+$.

\noindent
Specifically, we compare the ratio of rates for the decay chains 
$D^{*+} \rightarrow D^0 \pi^+, D^0 \rightarrow K^- e^+ \nu_e$,
 and $ D^0 \rightarrow K^- \pi^+$  in data and simulated events, 
this way, many systematic uncertainties cancel, 

{\small
\begin{eqnarray}
R_D &=& \frac{BR(D^0 \rightarrow K^- e^+ \nu_e)_{\rm data}}{BR(D^0 \rightarrow K^- \pi^+)_{\rm data}} =  
\frac{BR(D^0 \rightarrow K^- e^+ \nu_e)_{\rm MC}}{BR(D^0 \rightarrow K^- \pi^+)_{\rm MC}}\nonumber\\
 &\times & \frac{N(c\bar{c})_{Ke\nu}}{N(c\bar{c})_{K\pi}} \times
\frac{{\cal L}(data)_{K\pi}}{{\cal L}(data)_{Ke\nu}} \nonumber\\ 
 &\times & \frac{N(D^0 \rightarrow K^- e^+ \nu_e)_{\rm data}}{N(D^0 \rightarrow K^- e^+ \nu_e)_{\rm MC}} \times
\frac{N(D^0 \rightarrow K^- \pi^+)_{\rm MC}}{N(D^0 \rightarrow K^- \pi^+)_{\rm data}} \nonumber\\
 &\times & \frac{\epsilon(D^0 \rightarrow K^- e^+ \nu_e)_{\rm MC}}{\epsilon(D^0 \rightarrow K^- e^+ \nu_e)_{\rm data}} \times
\frac{\epsilon(D^0 \rightarrow K^- \pi^+)_{\rm data}}{\epsilon(D^0 \rightarrow K^- \pi^+)_{\rm MC}}
\label{eq:rd}
\end{eqnarray}
}

The first line in this expression is the ratio of the 
branching fraction for the two channels used in the simulation:
\beq
\frac{BR(D^0 \rightarrow K^- e^+ \nu_e)_{\rm MC}}{BR(D^0 \rightarrow K^- \pi^+)_{\rm MC}} = \frac{0.0364}{0.0383}.
\eeq

The second line is the ratio of the number of $c\bar{c}$
simulated events and the integrated luminosities for the
two channels:
\beq
\frac{N(c\bar{c})_{Ke\nu}}{N(c\bar{c})_{K\pi}} 
\frac{{\cal L}({\rm data})_{K\pi}}{{\cal L}({\rm data})_{Ke\nu}}=
\frac{117.0 \times 10^6}{117.3\times 10^6} \times \frac{73.43 \fb^{-1}}{74.27 \fb^{-1}}
\eeq

The third line corresponds to the ratios of measured numbers of 
signal events
in data and in simulation, and the last line gives the ratios of the efficiencies  
to data and simulation.

\subsection{Selection of candidate signal events} 

The selection of $D^0 \rightarrow K^- e^+ \nu_e$ candidates is explained in
section \ref{sec:Analysisq2}. For the rate measurement, the constraint on the $\Dstarp$ mass is not 
applied and also the momentum of the soft pion candidate is not included in 
the Fisher discriminant variable designed to suppress $B\bar{B}$-background.
Since generic simulated signal events used in this measurement have been generated with
the ISGW2 model, they have been weighted so that their
$q^2$ distribution agrees with the measurement presented in this paper.
Furthermore, we require for the Fisher discriminant $F_{B\bar{B}}>0$ and restrict 
$\delta(m) < 0.16~\GeVcd$.  
After background subtraction, there remain $76283 \pm 323$ and 
$95302 \pm 309$ events in data and simulation, respectively.
This gives:
\beq
\frac{N(D^0 \rightarrow K^- e^+ \nu_e)_{\rm data}}{N(D^0 \rightarrow K^- e^+ \nu_e)_{\rm MC}}=
0.8004 \pm 0.0043.
\eeq

\begin{figure*}[!htb]
  \begin{center}
    \mbox{\epsfig{file=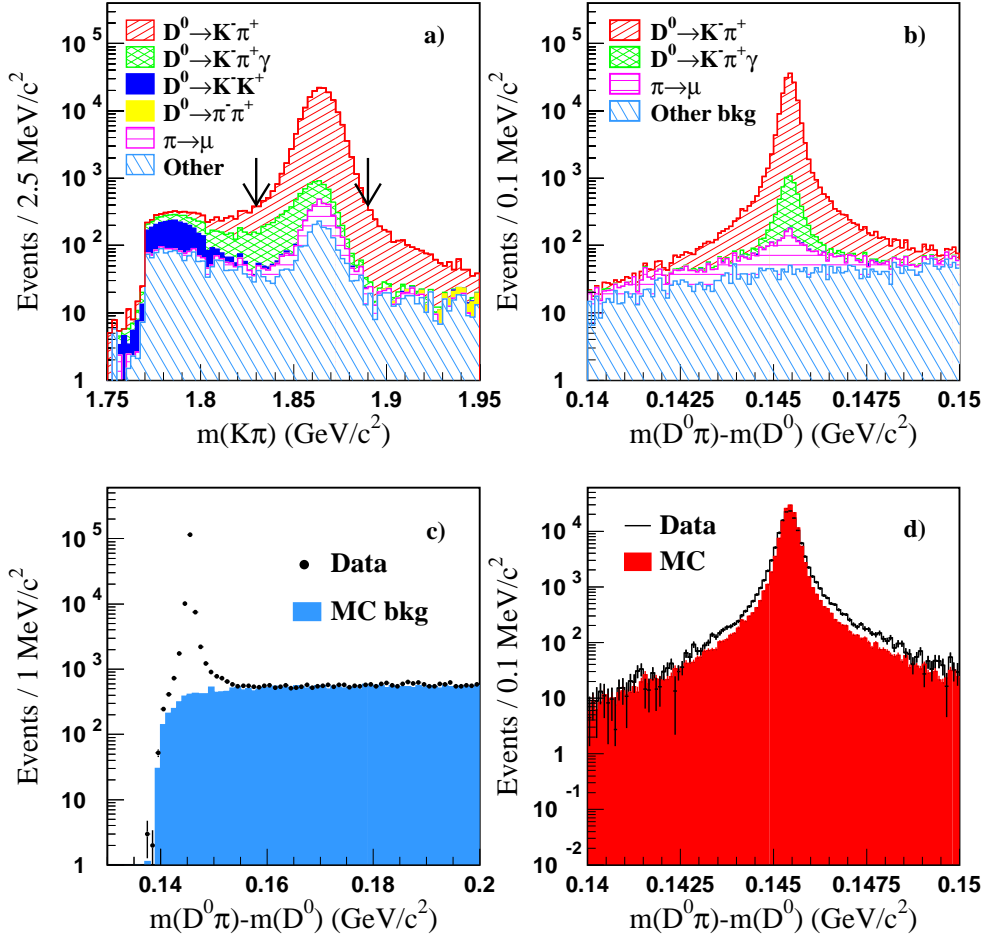,width=.80\textwidth,angle=0}}
  \end{center}
  \caption[]{ Events selected for the reference channel
$\Dstarp \rightarrow \Do \pi^+,~\Do \rightarrow \Km \pi^+$.
a) $K\pi$ mass distribution for events selected
in the range $\delta(m)\in [0.143,~0.148]~\GeVcd$.
b) $\delta(m)$  distribution for events selected
in the range $m(K\pi)\in [1.83,~1.89]~\GeVcd$. 
c) same distribution as in b), displayed on a larger mass range with
the non-peaking background indicated (shaded area).
d)  $\delta(m)$  distribution after non-peaking background
subtraction, data (points with statistical errors) and simulated events (shaded histogram).}
  \label{fig:dstarkpi2}
\end{figure*}

To select $D^0 \rightarrow K^- \pi^+ $ candidates,
the
same data samples
are used and particles, in each event, are selected in the same way. 
The same selection criteria on the Fisher discriminant to suppress $B\bar{B}$-events, on the
thrust axis direction and on other common variables are applied. 
Events are also analyzed in the same way, with
two hemispheres defined from
the thrust axis. 
In each hemisphere
a $D^0$ candidate is reconstructed by combining a charged $K$ with a pion
of opposite sign. These tracks have to form a vertex and the $K\pi$ mass 
must be within the range $[1.77,~1.95]~\GeVcd$. 
Another charged pion of appropriate sign 
is added to form a $D^{*+}$ candidate.  

In addition, the following selection criteria are used:
\begin{itemize}
\item the fraction of the beam momentum, in the c.m. frame, taken by the
$D^*$ candidate must exceed 0.48 to remove contributions
from $B\bar{B}$ events;
\item the measured $D^0$ mass must be
in the range between 1.83 and 1.89 $\GeVcd$. This requirement eliminates possible
contributions from remaining $D^0 \rightarrow K^-K^+$ or $\pi^+ \pi^-$
decays (see Fig.\ \ref{fig:dstarkpi2}-a);
\item the vertex fit  for the $D^0$ and $D^*$ have to converge.
\end{itemize}

The $\delta(m)$ distribution for candidate events 
is shown in Fig.\ \ref{fig:dstarkpi2}-c. 
The following components
contribute to the $D^{*+}$ signal (see Fig.\ \ref{fig:dstarkpi2}-b):
\begin{itemize}
\item $D^0 \rightarrow K^-\pi^+$ with no extra photon;
\item $D^0 \rightarrow K^-\pi^+$ with at least one extra photon;
\item $D^0 \rightarrow K^-\pi^+(\gamma)$ where the $\pi^+$, mainly from
 the $D^{*+}$, decays into a muon.
\end{itemize}

The $\delta(m)$ distribution corresponding to other
event categories does not show a peaking component
in the $D^{*+}$ signal region. 
The 
total background level, is normalized  
to data using events in the $\delta(m)$ interval between 0.165
and 0.200 $\GeVcd$ (see Fig.\ \ref{fig:dstarkpi2}-c). 
This global scaling factor is equal to 
$1.069\pm0.011$.
After background subtraction, the $\delta(m)$ distributions obtained
in data and simulation can be compared in Fig.\ \ref{fig:dstarkpi2}-d.
Since the $D^{*+}$ signal is narrower in the simulation, we 
use a mass window such that this difference has a small
effect on the measured number of $\Dstarp$ events in data and in the 
simulation. 
There are $166960\pm409$ and $134537\pm 374$ candidates 
selected in the interval 
$\delta(m)\in [0.142,~0.149]~\GeVcd$ for simulated and data events respectively. 
This means:
\beq
\frac{N(D^0 \rightarrow K^- \pi^+)_{\rm MC}}{N(D^0 \rightarrow K^- \pi^+)_{\rm data}}=
1.230 \pm 0.0046.
\eeq

\subsection{Efficiency corrections} 

The impact of the selection requirement on the reconstructed $K \pi$ mass, 
has been studied. The $K\pi$ mass distribution signal
for simulated events (not including radiative
photons) is compared with the corresponding distribution obtained
with data after background subtraction. The background
contributions are taken from the simulation.
The fraction of $D^0$ candidates in the selected mass range
(between 1.83 and 1.89 $\GeVcd$) is 
$(97.64 \pm 0.25)\%$ in MC and  $(97.13 \pm 0.29)\%$ in data events. 
The ratio of efficiencies is equal to:
\beq
\frac{\epsilon(D^0 \rightarrow K^- \pi^+)_{\rm data}}{\epsilon(D^0 \rightarrow K^- \pi^+)_{\rm MC}}=
0.9947 \pm 0.0039.
\eeq

\begin{figure}[!htb]
  \begin{center}
    \mbox{\epsfig{file=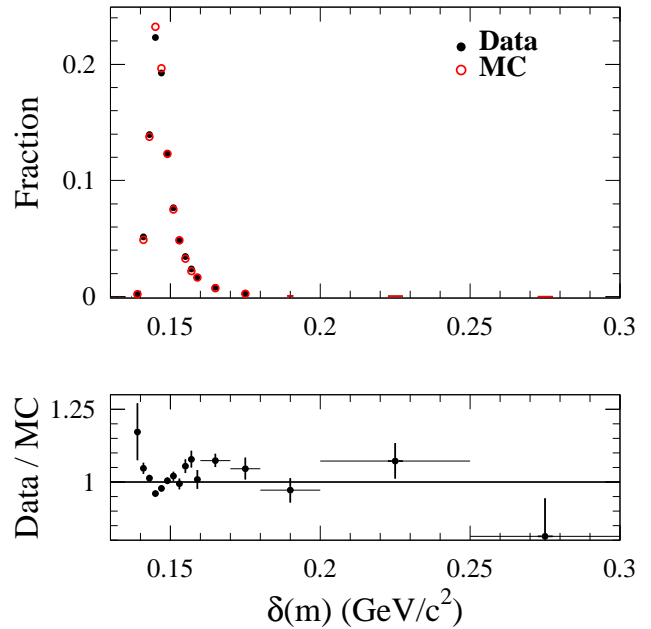,width=.54\textwidth,angle=0}}
  \end{center}
  \caption[]{$\delta(m)$ distribution for $\Do \rightarrow \Km \pi^+ \pi^0$
events analyzed as if they were semileptonic decays. Distributions have been
normalized to unity; note that the bin size is not uniform. The bottom plot 
shows the ratio of the two
distributions above.}
  \label{fig:dmkpipi0}
\end{figure}

Since $\Do \rightarrow K^- e^+ \nu_e$ events have been selected using a selection requirement
on $\delta(m)$, 
we need to confirm that the distribution of this variable is similar in
data and simulation. This is checked 
 by comparing the 
distributions obtained with $D^0 \rightarrow K^- \pi^+ \pi^0$
events analyzed as if they were
semileptonic decays. The $\delta(m)$ distributions are compared 
in Fig.\ \ref{fig:dmkpipi0}. Below
0.16 $\GeVcd$, there are $0.93552\pm0.00066$ of the $\Dstarp$ candidates
in the simulation and $0.93219\pm0.00078$ for data. The corresponding
ratio of efficiencies (MC/data) is equal to $1.0036\pm0.0010$.

Using $D^0 \rightarrow K^- \pi^+ \pi^0$
events, we also measure the  difference 
between the fraction of events retained 
after the mass-constrained fits.
Namely, it is $0.98038 \pm 0.00037$ in the simulation compared to 
$0.97438 \pm 0.00049$ in data. The relative efficiency (MC/data) for this selection 
is $1.0062 \pm 0.0006$.
Based on these two measured corrections the ratio of efficiencies are:
\beq
\frac{\epsilon(D^0 \rightarrow K^- e^+ \nu_e)_{\rm MC}}{\epsilon(D^0 \rightarrow K^- e^+ \nu_e)_{\rm data}}=
1.0098\pm 0.0011.
\eeq
The quoted uncertainties, in this section, are of statistical
origin and will be included in the statistical uncertainty on $R_D$.
Other differences between the two analyzed channels are considered in the
following section and contribute to systematic uncertainties.

\subsection{Systematic uncertainties on \boldmath{$R_D$}} 

A summary of the systematic uncertainties on $R_D$ are given in Table \ref{tab:systrate}.
They originate from selection criteria which 
are different for the two channels. Some of these uncertainties are the same
as those already considered for the determination of the $q^2$ variation
of $f_+$.

\subsubsection{Correlated systematic uncertainties}
Systematic uncertainties on the decay rate coming from effects that contribute
in the measurement of the $q^2$ dependence
of $f_+(q^2)$ are evaluated in Section \ref{sec:Systematics} and the
full covariance matrix for the measurements of the number of $D^0 \rightarrow K^- e^+ \nu_e$
signal events and the fraction of the decay spectrum fitted in each of the ten bins is
determined. 
Among the sources of systematic uncertainties, listed in Table \ref{tab:systall}, those corresponding to: 
\begin{itemize}  
\item the reconstruction algorithm,
\item the tuning of the resolution on $q^2$,
\item the corrections applied on electron identification, and
\item the background normalization
\end{itemize}
are taken as common sources. Corresponding relative uncertainties on $R_D$
are given in Table \ref{tab:systrate}.

Other systematic uncertainties contributing to the form factor
measurement also affect the reference channel and so their effects on $R_D$
cancel. They are related to the $c$-hadronization tuning and to the corrections 
applied on the kaon identification. 

\subsubsection{Selection requirement on the Fisher discriminant}
The stability of the fraction of $\Do \rightarrow K^- e^+ \nu_e$ 
events selected in data and 
in simulation 
as a function of the Fisher discriminant, $F_{cc}$, designed
to suppress $c\bar{c}$ background has been examined.
This is done by comparing the distributions
of this variable measured in data and in simulation
as given in Fig.\ \ref{fig:fisherkenu} for two selected intervals in $\delta(m)$.

The value corresponding to $F_{cc}>0$ and for events 
selected in the range $\delta(m)<0.16~\GeVcd$ is used as the central result and 
half the difference between the measurements corresponding
to $F_{cc}$ greater than $-0.25~{\rm and}~+0.25$ is taken as systematic uncertainty. This range corresponds
to a relative change of $40\%$ of the efficiency for signal events,
and gives an uncertainty of $\pm 0.0061$ on the ratio of
data and simulated signal candidates.

\subsubsection{$\Dstarp$ counting in $\Do \rightarrow K^- \pi^+$}
$\Dstarp$ candidates are selected in the range 
$\delta(m)\in [0.142,~0.149]~\GeVcd$. From the simulation it is expected
that the fraction of signal events outside this interval is equal
to 1.4$\%$. Even though the $\Dstarp$ signal is slightly narrower in the simulation,
there is not a large discrepancy in the tails.
The fraction of signal events measured in the sidebands $\delta(m)\in [0.140,~0.142]\oplus
[0.149,~0.150]~\GeVcd$ is $0.4\%$ and  $0.5\%$,
respectively, for simulation and data. An uncertainty of $\pm0.004$,
corresponding to 30$\%$ uncertainty on the total fraction of events
outside the selected $\delta(m)$ interval is assumed.

\begin{table}[!htb]
  \caption[]{ {Summary of systematic uncertainties on the relative decay 
rate measurement.}\label{tab:systrate}}
\begin{center}
  \begin{tabular}{lc}
    \hline\hline
Source & Relative variation \\
\hline
Reconstruction algorithm & $\pm 0.42\%$\\
Resolution on $q^2$ & $\sim 0$ \\
Electron ID & $\pm0.56\%$ \\
Background subtraction& $\pm0.63\%$\\
\hline
Cut on Fisher variable & $\pm 0.76\%$\\
$\Dstarp$ counting ($\Do \rightarrow K^- \pi^+$) & $\pm0.40\%$ \\
\hline
Total & $\pm1.27\%$ \\
\hline\hline
\end{tabular}
\end{center}
\end{table}

\subsection{Decay rate measurement}

Combining all measured fractions in  Eq. (\ref{eq:rd}), 
the measured relative decay rate is:

\beq
R_D = 0.9269 \pm 0.0072 \pm 0.0119.
\eeq
\noindent
Using the world average for the branching fraction
$BR(\Do \rightarrow \Km \pi^+) =(3.80 \pm 0.07) \%$
\cite{ref:pdg06}, gives
$BR(D^0 \rightarrow K^- e^+ \nu_e(\gamma)) = (3.522 \pm 0.027 \pm 0.045 \pm 0.065)\%$,
where the last quoted uncertainty corresponds to the 
accuracy on $BR(D^0 \rightarrow K^- \pi^+)$.

\section{Summary}
\label{sec:Summary}
The decay rate distribution for the channel
$\Do \rightarrow \Km e^+ \nu_e(\gamma)$ has been measured
in ten bins in Table \ref{tab:errmeas}. 
Several 
theoretical expectations for the variation of this form factor with $q^2$ 
have been considered and values for the corresponding parameters have been
obtained (see Table \ref{tab:fittedparam}).
The $q^2$ variation of the 
form factor can be parameterized with a single parameter using
different expressions. 
The ISGW2 model with expected values for the parameters is excluded,
as is the pole mass parameterization with $m_{\rm pole}=m_{D_s^*}$.

The value of the decay branching fraction has been also 
measured independent of a model.
 
Combining these measurements, the value of the hadronic form
factor is obtained:
\beq
f_+(0)=\frac{1}{\left | V_{cs} \right |}
\sqrt{\frac{24 \pi^3}{G_F^2}\frac{BR}{\tau_{D^0} I}},
\eeq
\noindent where
$BR$ is the measured $D^0 \rightarrow K^- e^+ \nu_e$ branching fraction, 
$\tau_{D^0}=(410.1\pm1.5)\times10^{-15}~s$ \cite{ref:pdg06} is the $D^0$ lifetime and
$I=\int_0^{q^2_{\rm max}}{\left | \vec{p}_K (q^2) \right |^3 \left |f_+(q^2)/f_+(0) \right |^2}~dq^2$.
To account for the variation of the form factor
within one bin, and in particular to extrapolate the result at $q^2=0$, the pole
mass and the modified pole ansatze have been used; the corresponding values obtained for 
$f_+(0)$ differ by 0.002. Taking the average between these two values and including
their difference in the systematic uncertainty,
this gives
\beq
f_+(0) = 0.727 \pm 0.007 \pm 0.005 \pm 0.007, 
\eeq
where the last quoted uncertainty corresponds to the accuracy
on $BR(D^0 \rightarrow K^- \pi^+)$, $\tau_{D^0}$ and $\left | V_{cs} \right |$.
It agrees with expectations and in particular with LQCD computations \cite{ref:lqcd3}.
Using the $z$ expansion of Eq. (\ref{eq:taylor}), we find 
$a_0=(2.98 \pm 0.01 \pm0.03 \pm0.03)\times 10^{-2}$.

The high accuracy of the present measurement will be a reference test
for improved lattice determinations of the $q^2$ variation of $f_+$.

\vspace{0.5cm}
\section{Acknowledgments}
\label{sec:Acknowledgments}

The authors wish to thank R. J. Hill, D. Becirevic, C. Bernard,
Ph. Boucaud, S. Descotes-Genon, L. Lellouch, J.-P. Leroy, A. Le Yaouanc and O. P\`ene for their 
help with the theoretical interpretation of these results.

We are grateful for the 
extraordinary contributions of our \pep2\ colleagues in
achieving the excellent luminosity and machine conditions
that have made this work possible.
The success of this project also relies critically on the 
expertise and dedication of the computing organizations that 
support \babar.
The collaborating institutions wish to thank 
SLAC for its support and the kind hospitality extended to them. 
This work is supported by the
US Department of Energy
and National Science Foundation, the
Natural Sciences and Engineering Research Council (Canada),
Institute of High Energy Physics (China), the
Commissariat \`a l'Energie Atomique and
Institut National de Physique Nucl\'eaire et de Physique des Particules
(France), the
Bundesministerium f\"ur Bildung und Forschung and
Deutsche Forschungsgemeinschaft
(Germany), the
Istituto Nazionale di Fisica Nucleare (Italy),
the Foundation for Fundamental Research on Matter (The Netherlands),
the Research Council of Norway, the
Ministry of Science and Technology of the Russian Federation, 
Ministerio de Educaci\'on y Ciencia (Spain), and the
Particle Physics and Astronomy Research Council (United Kingdom). 
Individuals have received support from 
the Marie-Curie IEF program (European Union) and
the A. P. Sloan Foundation.


\begin{thebibliography}{99}
\bibitem{ref:chargeconj}
Charge conjugate states are implied throughout this analysis.
\bibitem{ref:chiu}
The variation of $\chi_V$ with $u=m_s/m_c$ is small, since
$\chi_V(0.33)=1.02 \times \chi_V(0)$.
\end{thebibliography}
\end{document}